\documentclass[journal,compsoc]{IEEEtran}
\usepackage{multirow} 
\usepackage[pdftex]{graphicx}
\usepackage{amsmath}
\usepackage[cmintegrals]{newtxmath}

\usepackage{colortbl}
\usepackage{hhline}
\usepackage{array}

\usepackage{cite}
  \usepackage[pdftex]{graphicx}

\begin{document}
%
\title{Towards 5G Enabled Tactile Robotic Telesurgery}
%
%
%
%

\author{Qi Zhang,
        Jianhui Liu,
        and Guodong Zhao
\IEEEcompsocitemizethanks{\IEEEcompsocthanksitem Qi Zhang and Jianhui Liu are with the Department
of Engineering, Aarhus University, Aarhus N,
8200 Denmark. E-mail: qz@eng.au.dk and jianhui.liu@eng.au.dk 
\IEEEcompsocthanksitem Guodong Zhao is with University of Electronic Science and Technology of
China, Chengdu, China. E-mail: gdzhao@uestc.edu.cn }}
\maketitle
\begin{abstract}

Robotic telesurgery has a potential to provide extreme and urgent health care services and bring unprecedented opportunities to deliver highly specialized skills globally. It has a significant societal impact and is regarded as one of the appealing use cases of Tactile Internet and 5G applications. However, the performance of robotic telesurgery largely depends on the network performance in terms of latency, jitter and packet loss, especially when telesurgical system is equipped with haptic feedback. This imposes significant challenges to design a reliable and secure but cost-effective communication solution. This article aims to give a better understanding of the characteristics of robotic telesurgical system, and the limiting factors, the possible telesurgery services and the communication quality of service (QoS) requirements of the multi-modal sensory data. Based on this, a viable network architecture enabled by the converged edge and core cloud is presented and the relevant research challenges, open issues and enabling technologies in the 5G communication system are discussed.

\end{abstract}


\IEEEpeerreviewmaketitle


\section{Introduction}\label{sec:introduction}

Tactile Internet coined by Prof. Fettweis envisions communication network is moving from communicating data, voice, and video to real-time steering and control. Indeed, Tactile Internet will create a new dimension for human-machine interaction. Tactile Internet has been regarded as the next wave of innovation after Internet of Things, with an ambition to enable "Internet of skills" globally. 
As one of the appealing use cases, robotic telesurgery has attracted tremendous interest from industry and academia. NASA has been working on to send surgical robots to space station. Many big network industrial players, e.g., Vodafone, Huawei, Nokia Bell labs, etc. are developing communication solutions for telesurgery services. Ericsson and King's College London demonstrated telesurgery in 5G world event 2016. With the advancement in next-generation wireless communication and networking technologies, it will no long be in fiction that surgeons perform operations for patient remotely in space, in battle field, or in a moving ambulance. 

\textbf{Motivations} of this article are to give an introduction to the characteristics of telesurgical system and its QoS requirements. Through a better understanding of telesurgical system, we shed light on the communication and networking research directions for telesurgery. This article aims to inspire researchers to design reliable and secure but cost-effective communication solutions tailored for tactile robotic telesurgery.

Robotic telesurgery, i.e., remote surgery, envisions the surgeon and patient are geographically separated. Telesurgery creates a new way to provide extreme and urgent health care. A telesurgical system will be developed based on robotic surgical system which consists of master console and slave robot, i.e., teleoperator. Master console is a human-system interface (HSI) which is composed of a haptic device for position-orientation input, a video display and headphones for video and voice feedback and in the future it will have haptic feedback output. The teleoperator is equipped with 3D video camera, a microphone and will have a number of force sensors and tactile sensors in the future. The da Vinci Surgical System is currently the most widely used robotic surgical system and provides visual sensory only. In telesurgergy, the master console and teleoperator will be connected by reliable high speed communication network to transport the real-time manipulation commands and multi-modal sensory data. 

Besides the high cost, the major limitation factor hinders the current robotic system to become the standard technique of minimally invasive surgery worldwide is the lack of effective haptic feedback including force (kinesthetic) and tactile (cutaneous) feedback~\cite{Ghezzi2016}. Force feedback can enhance the dexterity and controllability of surgical instrument thereby improving the surgical precision particularly for those difficult manipulations e.g.~coronary artery bypass grafting and reducing potential unintentional injuries during a dissection task~\cite{Konstantinova2014}. 
Tactile sensing can determine mechanical properties of tissue e.g.~viscosity and surface texture, which helps to detect exact tumor and lump locations in organ, allowing surgeon to accurately identify the margin of the abnormality to better conduct dissection~\cite{Konstantinova2014}.

In 2001, the first transaltantic telesurgery was demonstrated on the patient in Strasbourg, France by the surgeon located in New York. It used a dedicated asynchronous transfer mode (ATM) fiber optic communication link with a constant latency 155~ms and no packet loss~\cite{Marescaux2002}. Despite the successful demonstration, telesurgery has not been widely used. The main reason lies in high communication cost, long latency, no reliability guarantee by public Internet. Some research work has studied the impact of latency on telesurgical performance using robotic simulator dV-Trainer based on visual feedback only~\cite{Perez2015ImpactOD}. The observations are that the surgical performance deteriorates as latency increases. 
The latency impact is related to the surgical procedure complexity and the skills of individual surgeon. 
The delay impact is mild below 200~ms. 
Surgeons are able to adapt to the delay through training under a constant delay. However, it is challenging to conduct telesurgery with variable latency. 

With direct haptic feedback, the tolerable latency and jitter will become more stringent, as small communication latency, jitter or packet loss will cause instability in the closed haptic control loop. It is necessary to limit the synchronization error to prevent cyber sickness~\cite{Simsek2016, Aijaz2016J}. Hence, the challenge is not only about minimizing the latency and jitter, but also synchronization between the 3D visual signal and haptic feedback.

\textbf{Main contribution} of this article: (i) identify the limiting factors for telesurgical system, (ii) categorize telesurgery services and give the roadmap of their development, (iii) present a viable network architecture, (iv) summarize relevant research challenges and open issues, (v) discuss how the enabling technologies in 5G can address these challenges.

 
\section{Communication QoS Requirement for Robotic Telesurgery}\label{Sec:CommQoS}
\begin{figure}[!t]
	\centering
	\includegraphics[width=3.5in]{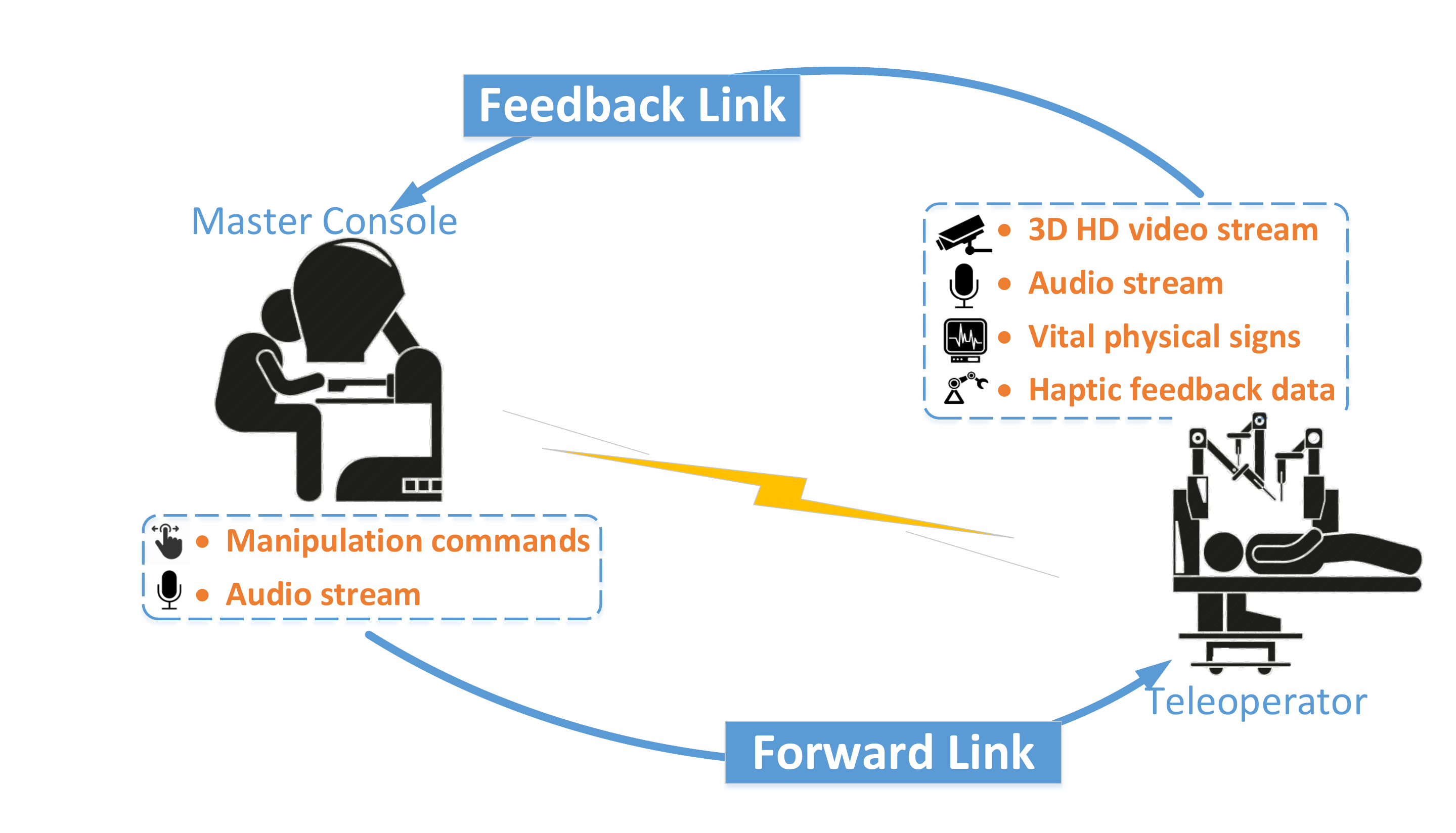}
	\caption{A close control loop between master console and teleoperator.}
	\label{fig_data}
\end{figure}

The concept of master-slave robotic surgical system theoretically allows the realization of telesurgery. In the da Vinci system a proprietary short-distance communication protocol over optic fiber connects the master console and teleoperator. The telesurgical system is a close-loop communication system consisting of forward link and feedback link, as shown in Fig.~\ref{fig_data}. 

\textbf{Forward link} transports real-time manipulation commands to control the motions and rotation of robotic arms at the teleoprator, along with voice stream from the surgeon to communicate with the surgical team remotely.

\textbf{Feedback link} transports real-time multi-modal sensory feedback from the teleoperator, including 3D video stream, force feedback e.g., pressure, and tactile feedback e.g. tissue mechanical properties, and patient's physiological data e.g. blood pressure and heart rate, along with voice stream from assistant nurses, anaesthetist and other collaborating surgeons at the patient's side. 

The communication QoS requirements of multi-modal sensory data in telesurgery are listed in Table~\ref{table_requirement} and the details are elaborated below.

\begin{table*}
	\definecolor{lightblue}{rgb}{0.93,0.93,0.94}
	\definecolor{deepblue}{rgb}{0.80,0.88,0.92}
	\newcolumntype{a}{>{\columncolor{deepblue}}c}
	\newcolumntype{b}{>{\columncolor{lightblue}}c}
	\arrayrulecolor[gray]{0.6}
	\renewcommand{\arraystretch}{2.3}
	\setlength\arrayrulewidth{0.8pt}
	\newlength\savedwidth
	\newcommand\whline{\noalign{\global\savedwidth\arrayrulewidth
			\global\arrayrulewidth 0.8pt}%
		\hline
		\noalign{\global\arrayrulewidth\savedwidth}}
	\caption{Communication QoS requirements for the multi-modality sensory data in robotic telesurgery}
	\label{table_requirement}
	\centering
	\begin{tabular}{a b a b a b a}
		& \textbf{Data types} & \textbf{Latency} & \textbf{Jitter} & \textbf{Packet Loss Rate} & \textbf{Data Rate} & \textbf{Ref}\\
		\arrayrulecolor{black}\whline
		& 2D camera flow & $< 150$ ms & 3-30 ms &  $< 10^{-3}$ & $< 10 $ Mbps & \cite{Perez2015ImpactOD, Admux}\\
		\hhline{|*1{>{\arrayrulecolor{deepblue}}-}*6{>{\arrayrulecolor[gray]{0.6}}-}|}
		& 3D camera flow & $< 150$ ms & 3-30 ms & $< 10^{-3}$ & 137 Mbps - 1.6Gbps &\cite{Perez2015ImpactOD, Admux }\\
		\hhline{|*1{>{\arrayrulecolor{deepblue}}-}*6{>{\arrayrulecolor[gray]{0.6}}-}|}
		\multirow{-3}{*}{\textbf{Real-time multimedia stream}} & Audio flow & $<150$ ms & $<30$ ms & $< 10^{-2}$ & $22-200$ Kbps & \cite{Cizmecis2017, Admux}\\
		\hline
		& Temperature & $< 250$ ms & - &$< 10^{-3}$ & $< 10$ kbps & \cite{Patel2010}\\
		\hhline{|*1{>{\arrayrulecolor{deepblue}}-}*6{>{\arrayrulecolor[gray]{0.6}}-}|}
		& Blood pressure & $< 250$ ms & - & $<  10^{-3}$ & $< 10$ kbps & \cite{Patel2010}\\
		\hhline{|*1{>{\arrayrulecolor{deepblue}}-}*6{>{\arrayrulecolor[gray]{0.6}}-}|}
		& Heart rate & $< 250$ ms & - & $< 10^{-3}$ & $< 10$ kbps & \cite{Patel2010}\\
		\hhline{|*1{>{\arrayrulecolor{deepblue}}-}*6{>{\arrayrulecolor[gray]{0.6}}-}|}
		& Respiration rate & $< 250$ ms & - & $< 10^{-3}$ & $< 10$ Kbps & \cite{Patel2010}\\
		\hhline{|*1{>{\arrayrulecolor{deepblue}}-}*6{>{\arrayrulecolor[gray]{0.6}}-}|}
		& ECG & $< 250$ ms &- &$< 10^{-3}$ & $72$ kbps & \cite{Patel2010}\\
		\hhline{|*1{>{\arrayrulecolor{deepblue}}-}*6{>{\arrayrulecolor[gray]{0.6}}-}|}
		& EEG & $< 250$ ms &  -&$< 10^{-3}$ & $86.4$ kbps & \cite{Patel2010}\\
		\hhline{|*1{>{\arrayrulecolor{deepblue}}-}*6{>{\arrayrulecolor[gray]{0.6}}-}|}
		\multirow{-7}{*}{\textbf{Physical vital signs}} & EMG & $< 250$ ms & - &$< 10^{-3}$ & $1.536$ Mbps & \cite{Patel2010}\\
		\hline
		& Force & $3-10$ ms & $<2$ ms & $< 10^{-4}$ & $128- 400$ Kbps & \cite{Admux,Cizmecis2017}\\
		\hhline{|*1{>{\arrayrulecolor{deepblue}}-}*6{>{\arrayrulecolor[gray]{0.6}}-}|}
		\multirow{-2}{*}{\textbf{Haptic feedback}} & Vibration & $< 5.5$ ms & $<2$ ms & $< 10^{-4}$ & $128- 400$ Kbps  & \cite{Cizmecis2017, Admux,Vibration }\\
	\end{tabular}
\end{table*}

For 3D video visualization, two HD video streams from surgical system are processed and displayed as a single data stream to the 3D display device at the required resolution, frame rate and encoding standard. For 3D MediVision products, the resolution of 1920x1080 pixels with 32bits/pixel and frame rate of 30 frame per second (fps) or 60fps per channel, i.e., 60fps or 120fps for 3D display device. Assuming the medical video compression rate in the range of 1:29 to 1:5, the required data rate is between 137 Mbps to 1.6 Gbps. The delay should be below 150~ms to ensure operation performance~\cite{Perez2015ImpactOD}. No comprehensive experiments have tested the tolerable jitter and packet loss for telesurgery. Video conferencing requires jitter below 30~ms and packet loss rate below 1\%~\cite{Admux}. 

For haptic data transmission, the packet payload is small which depending on the number of degrees of freedom and the sample resolution. Therefore, the overhead of communication protocol header becomes not trivial and should be included in data rate estimation. 
The typical sampling rate of uncompressed haptic signal is 1kHz, i.e. 1ms transmission service time per sample. Perception-based haptic data reduction according to Weber's Law is feasible. 
It is possible to achieve average sample rate reduction up to 90\% with satisfactory subjective ratings~\cite{Cizmecis2017}. 
For general teleoperation system not specific for telesurgery~\cite{Admux, Cizmecis2017}, the data rate for haptic data varies from 128 to 400 kbps, the latency and jitter should be below 50~ms and 2~ms, respectively, and the required data loss is about 0.01\%. 
The study in \cite{Vibration} investigated the effect of vibration feedback latency on material perception and found the latency threshold is 5.5~ms. Longer latency gives the user a wrong sensation of the mechanical properties. Telesurgery is mission critical and requires high precision in manipulation, therefore, more stringent latency, jitter and packet loss are needed.  

The physiological data can tolerate 250~ms latency with moderate data rates~\cite{Patel2010}. The packet loss in telesurgery is not as critical as in body area network, as it is the local anesthetist responsible for monitoring the physical vital signs. The remote surgeon mainly relies on the 3D video stream and haptic feedback.

For voice conversation, the data rate ranges from 22 to 220~kbps. The delay and jitter should be smaller than 150~ms and 30~ms, respectively. It can tolerate 1\% packet loss~\cite{Admux}.

\section{Envisioned Telesurgery Services and Network Architecture}
In this section we briefly describe the possible telesurgery services and explain a network architecture with the involved communication and networking protocols.

\subsection{Envisioned Telesurgery Services}
\begin{figure}
	\centering
	\includegraphics[width=3.5in]{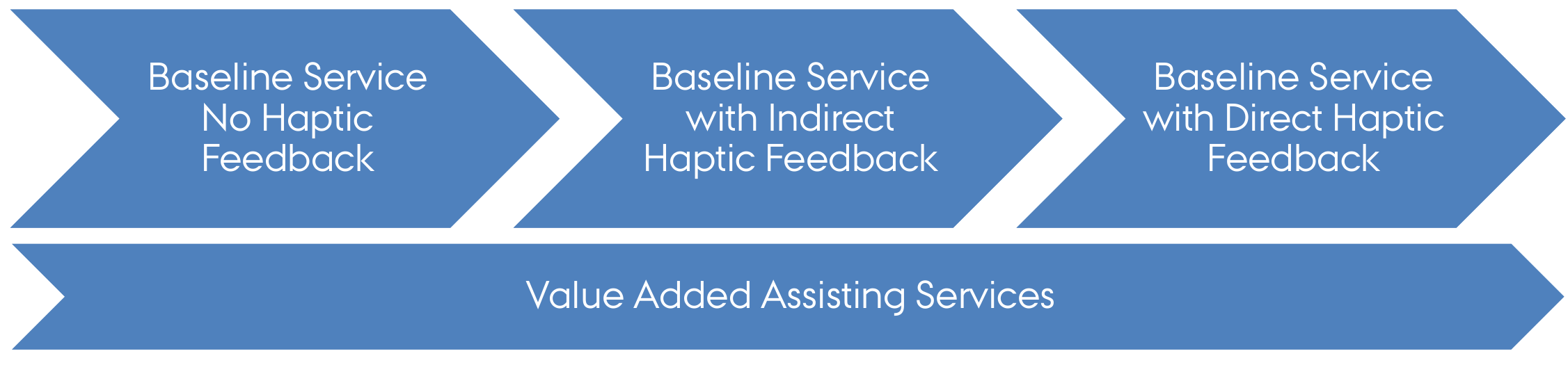}
	\caption{The roadmap of robotic telesurgery services development.}
	\label{fig_TelesurgeryRoadmap}
\end{figure}

Telesurgery services can basically be categorized into baseline services and value added assisting services as shown in Fig.~\ref{fig_TelesurgeryRoadmap}.

\textbf{Baseline telesurgery services} will be developed through several stages. (i) No haptic feedback: surgeon currently operates the master console to manipulate the teleoperator solely resorting to the 3D video stream \cite{Marescaux2002, Perez2015ImpactOD}. (ii) With graphical display of haptic feedback: It is an indirect haptic feedback as a sensory substitution system and not intuitive, therefore, requires a longer learning curve \cite{Okamura2009}. (iii) With direct haptic feedback: the robotic arms will be equipped haptic sensors. Such systems are expected to improve the operation performance, especially for novices, as it can help surgeon make correct motions and shorten the learning curve\cite{Simsek2016}. The indirect or direct haptic feedback can also facilitate remote diagnosis.

\textbf{Value added assisting services} will be enabled by the advanced artificial intelligence (AI). The real-time multi-modal sensory data, pre-cached electronic medical records and related historical surgical records can be processed by data analytics and machine learning algorithms in the edge and core cloud. It can provide diverse assisting services to help surgeon in abnormality diagnosis and decision making. For instance, Verb Surgical's is going to leverage advanced imaging and artificial neural networks to process the 3D video and images and annotate the feed with anatomical data. It will help surgeons to interpret what they see and can even give information about the boundaries of a tumor and suggest dissection strategy. 

\subsection{Network Architecture}
To enable the baseline and value added assisting services for telesurgery and to meet the required QoS of multi-modal sensory data, the communication network shall at least support both enhanced mobile broadband (eMBB) to support 3D video and ultra-reliable low latency communication (uRLLC) for haptic feedback. The master console and teleoperator are connected through Internet. The access network can be 4G or 5G wireless networks or fixed broadband networks. Since mobile communication offers flexibility and has advantages in extreme healthcare scenarios e.g. in a moving ambulance, this article focuses on how the 5G communication system will enable tactile robotic telesurgery. 

\begin{figure}[h]
\centering
\includegraphics[width=3.5in]{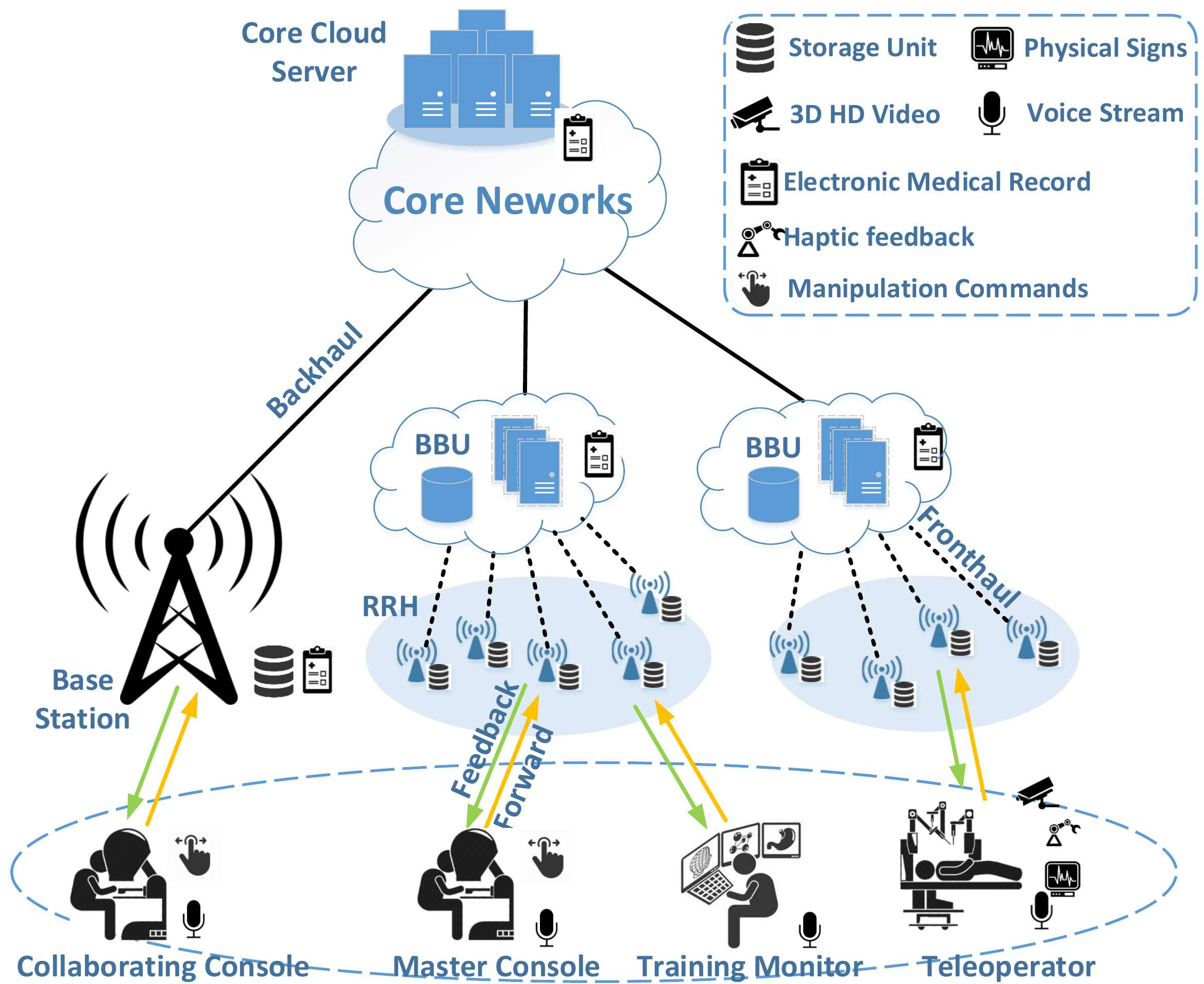}
\caption{A converged edge-cloud enabled network architecture.}
\label{fig_architecture}
\end{figure}

Based on the 5G communication system, a viable network architecture converging the edge and core cloud is illustrated in Fig.~\ref{fig_architecture}. The master console and teleoperator access the network through  distributed radio access network (D-RAN) or Cloud-RAN (C-RAN). Two geographically separated surgeons can collaborate in one surgery, i.e., manipulating the teleoperator remotely using a dual console which is available in the da Vinci SI Surgical System. It is also feasible to display the 3D video feedback at a third location for telesurgery demonstration, training and clinical guidance purposes.

The source-terminals are responsible for encoding sensory data. Different compression schemes for multi-modal sensory data are needed to reduce the data rate, e.g., H.264 for video, and perception-based data reduction for haptic feedback. To be robust against transmission rate variation, dynamic codec configuration and scalable video coding can be implemented. Multiplexer can be applied to interleave different sensory packets into one serialized bitstream. To improve user experienced QoS, various control methods e.g., delay compensation, jitter smoothing and stability control, are needed at the end-terminal. 

D-RAN or C-RAN contains protocol stack to perform radio access functions, from radio frequency, physical layer, medium access control, radio link control, packet data convergence protocol to radio resource control. 
The functions like multiple access, radio resource allocation, Hybrid automatic repeat request (HARQ), etc. are implemented here. To support higher data rate and diverse services for the use cases from different verticals with reduced CAPEX and OPEX, C-RAN will mostly be used. In C-RAN, base band processing unit (BBU) is decoupled from remote radio head (RRH). RRH is located closer to the users able to serve higher data rate and lower latency traffic. BBUs pool is realized by a centralized data center which leverages cloud technology to dynamically and flexibly pool computational resource to perform baseband processing. Multiple RRHs are connected to the BBUs pool via high bandwidth and low latency fronthaul. Centralization in C-RAN offers multiplexing gain, inter-cell interference coordination, and energy savings. To relax the requirement on franthaul, flexible functional split of baseband functionality between RRH and BBU has been considered, i.e., keeping a subset of baseband functionality in RRH. Functional split is a trade-off between cell coordination gain and reducing fronthaul bandwidth. For telesurgery, the MAC-PHY split or low PHY split option could be suitable depending on the fronthaul quality.

The backhaul network based on MPLS (Multiprotocol Label Switching) or optical network are currently used to connect RAN and mobile core network. SDN (Software-defined networking) is expected to be applied in the backhaul network. The network elements in MPLS and optical network will be implemented by programmable hardware-based switches and routers or even virtualized switches which form the data plane and are controlled by the SDN controller e.g., Floodlight controller or OpenDaylight controller. In the backhaul network, functions like routing, network monitoring, congestion control and traffic-aware reconfiguration communicate with the SDN controller via Northbound API~\cite{Nguyen2016}.
  
The SDN-based mobile core network consists of simple network gateways through which it is connected to the Internet. The centralized SDN core controller can control QoS in a fine grained manner. Considering the differentiated QoS requirements, video packets and haptic data could take different forwarding paths. To increase reliability, multi-path routing can be used.

The data centers for BBUs pool and core cloud will pre-cache historic medical data, e.g., the electronic medical records and relevant surgical records. The AI engines located at edge and core cloud will actively participate in data analytics and machine learning for value added services, as well as perform radio and network analytics to predict network performance change (see Section~\ref{subsec:MEC}, \ref{subsec:NS} and \ref{subsec:AI}).

\section{Communication Research Challenges and Open Issues of Robotic Telesurgery}

Telesurgery imposes many critical research challenges including precision of robotic arms, bio-compatibility of robotic arms, and others. We focus on some of the most important communication research challenges and relevant open issues in this section.

\subsection{Multi-modal Sensory Information}

As shown in Fig.~\ref{fig_data}, multi-modal sensory data need to be transported from the teleoperator to master console. Different sensory modalities, i.e., visual, auditory and haptic, have different requirement in terms of sampling, transmission rate, latency, reliability, etc.~\cite{Aijaz2016J}. It is shown in Table~\ref{table_requirement}. How to synchronize between video stream and haptic feedback is challenging. Many researchers have worked on multiplexing scheme~\cite{Admux, Cizmecis2017}. The state-of-art multiplexing transmission schemes for teleoperation system can be designed at the transport or application layer. The majority of the research assumes a constant bit rate link with certain transmission capacity. 

The transport-layer approaches are developed on top of TCP or UDP protocols to optimize network scheduling based on weighted priorities~\cite{Cizmecis2017}. The application-layer approaches focus on efficient encoding and adaptive statistical multiplexing~\cite{Admux}. The difficulty is that large video packets can block haptic packets resulting delay and jitter violation. Although theoretically preemptive-assume scheduling strategy can prioritize haptic packets; it is not efficient in practice due to protocol header overhead to transmit small haptic packets at high packet rate~\cite{Cizmecis2017}. Currently multiplexer is not resilient to packet losses. Details about multiplexing schemes refer to~\cite{Cizmecis2017} and the references therein.

From radio resource management and information theory point of view, multiplexing might not be optimal to utilize the resource and achieve the capacity. It is an open question how to design resource management for uRLLC traffic (e.g., haptic data) when coexisting eMBB services (e.g., 3D video stream), since they are different in packet size and QoS requirements particularly latency and reliability. Moreover, video stream and haptic data need different error control schemes and congestion control methods.

\subsection{Ultra-low Latency and High Reliability}
The performance of telesurgery largely depends on the network performance, as latency, jitter and packet loss will jeopardize stability in the control system and degrade the performance of operation and system transparency~\cite{Cizmecis2017}. To ensure a stable control loop to complete high precision teleoperation tasks, e.g., suture, haptic data requires below 10~ms latency and $10^{-4}$ packet loss. Reliability and latency to some extent are interconnected. Degraded reliability can result in increased latency if using HARQ based scheme. For wireless communication system, communication latency and jitter can be caused by limited transmission bandwidth, congestion and uncertainty in transmission e.g. interference and transmission rate variation. 

Furthermore, Shannon capacity with arbitrary small error probability assumes infinite blocklength which is only relevant for large packet e.g. video packets. To ensure the reliability for short haptic packet with stringent delay, the maximal achievable rate with finite blocklength cannot be characterized by Shannon capacity. It is an open question and new information-theoretic metrics for finite packet length are needed.

\subsection{Security and Privacy}
When telesurgical system communicates over public network, it is exposed to various security and privacy issues which can have serious impact on the operation performance and even risks to threaten patient's life. 

For instance, malware programs can exploit the vulnerability in the robot system software to perform remote hijacking. Under the influence of malware programs, any unauthorized tasks, e.g. wrong dissection, can be performed by the remote malicious users (adversary). In Denial-of-Service (DoS) attacks, adversary can send a large number of flooding packets or any other spurious packets to the destination to consume excessive amounts of network bandwidth thereby resulting in packet delay, jitter and loss~\cite{SecTelesurgery}. DoS attack can shut off the telesurgical system from the Internet or dramatically slowing down network links hindering the communication between master console and teleoperator. Man-in-the middle attacks and impersonation attack could alter the manipulation commands and sensory feedback data. Nevertheless, it might not be easy for the adversary to change the video and haptic data consistently in real time. Additionally, there exists the typical privacy issue to exchange the patient's health data over Internet.

Therefore, security schemes should be implemented for access control, authentication, confidentiality and data integrity. Although security for communication network and cyber physical system has been advanced, it is challenging to simultaneously provide security and fulfill the delay and jitter requirement in telesurgery, as cryptography and security protocols will consume computational resource and introduce communication overhead. In fact, security support generally causes delay violation or communication QoS degradation~\cite{SecTelesurgery}. It is an open issue how to design cryptography and security schemes for real-time teleoperation system to meet the QoS and security constraints~\cite{SecTelesurgery}.

\subsection{High Cost of Communication Services}
Telesurgery has not become a medical routine basis. One of the limiting factors is high cost of the communication service. 
For the transatlantic telesurgery case, the connection between New York and European Institution of Telesurgery at Strasbourg was established through an ATM private network and 10~Mbps was reserved~\cite{Marescaux2002}. The cost for a 1-year availability of ATM lines point-to-point was between $ \$ 100,000$ and $\$ 200,000$ in 2001~\cite{Marescaux2002}. To enable telesurgery services to be widely used, cost-effective communication solutions must be in place.

\section{Enabling Technologies in 5G for Robotic Telesurgery}
In this section we highlight several key enabling technologies in 5G, although many other technologies can address the challenges in telesurgery, e.g. faster processor for video coding, time-delay compensation control architecture~\cite{Cizmecis2017}, biological inspiration of tactile sensing~\cite{Konstantinova2014}, physical unclonable function for low-cost authentication.

\subsection{Physical Layer Design for Ultra-Low Latency}
The design of 5G new radio (NR) air interface aims at 30-50X latency reduction in the RAN and several PHY designs have been proposed or specified in the standard. LTE has defined fast uplink access in Rel-14 which allows user to have a periodic uplink grant every TTI (time transmission interval). Fast uplink access avoids the delay waiting for grant. Sparse code multiple access based grant-free transmission for 5G in user centric no cell framework has demonstrated 94.9\% latency reduction~\cite{PoC2017}. 
Instead of using 1~ms TTI, a shorter TTI allows transmission duration smaller than one subframe, which has been specified in Rel-15. In the downlink, the data part of subframe can be split into several parts and each part can be scheduled independently by a new in-band channel. The OFDM symbol length in LTE and 5G NR is 66.67~$\mu$s and 16.67~$\mu$s, respectively. The shorter TTI or subframe 
enables short scheduling unit which can help to reach the goal of 5G user plane latency of 0.5~ms for uRLLC. Flexible and self-contained TDD subframe enables data transmission and its acknowledgement post-decoding to be contained within the same subframe. With the shorter slots and fewer HARQ interlace, it accelerates the associate control and feedback, thereby achieves lower latency.   

More elaboration about PHY design for ultra-low latency including OFDM numerology and frame structure is presented in~\cite{Simsek2016} and the references therein.

\subsection{Mobile Edge Computing}\label{subsec:MEC}

To provide value added assisting services for telesurgery, AI engines need to process large volume of real-time 3D video and images and tremendous pre-cached historic medical data, which requires large processing and storage power. 
To access traditional cloud infrastructure by Internet causes huge pressure to the backhaul and core network and long latency, which makes it simply not feasible to fulfil the latency requirement. 
Mobile edge computing (MEC) is a natural development to move the computing and storage unit to the edge of the network close to where data is generated. MEC is a new technology recognized as one of the key emerging technologies for 5G networks and the standards of MEC are being developed by ETSI~\cite{ETSIMEC}. With its unique capability, MEC is expected to improve latency and reliability, optimize network operation and improve user experience; particularly, multiple edge nodes are utilized simultaneously.  

MEC server can be deployed at multiple locations, e.g. LTE macro base station (eNodeB) site, RRH and BBUs pool of C-RAN, as they are already equipped with processing and storage unit. Thus most of the processing can be done within RAN and alleviate the bandwidth pressure to core network thereby shortening the latency. For example, pre-cached electronic medical records and relevant historic surgery data such as the recorded motion of robot arms can be stored in the data center at the BBUs pool. Leveraging the AI engine based on deep learning and pattern recognition algorithms, the MEC can process the 3D video and images to highlight the relevant features using scalable video coding and annotate the images of the diagnosis results, consequently minimize the required communication resource. One open question here is how to strike a balance between computation latency and communication latency. 

Additionally, radio analytics application in MEC can better estimate the channel quality and delay variation in the network. This information can be exploited to optimize the application level coding, channel coding and reduce video-stall occurrence~\cite{ETSIMEC} and facilitate flexible resource allocation. It is worth mentioning at due to the infrastructure and operation cost, the amount of processing power and storage at MEC is orders of magnitude smaller than the traditional clouds. Therefore, computing task split, offloading and content caching in MEC are the open research topics.

\subsection{Network Slicing}\label{subsec:NS} 
The 5G communication system aims to use a common communication infrastructure to support various use cases of different verticals with diverse and extreme requirements for latency, throughput, reliability and availability. Network slicing is an effective way to optimize 5G network to offer solutions tailored to all use cases by leveraging cloud technology, SDN, network function virtualization (NFV), end-to-end orchestration and management, and network analytics. It can address the latency, multi-modal sensory data, reliability, security and cost challenges in telesurgery.

Basically, network slice (NS) is composed of a set of network element functions with flexible granularity across from radio access, transport to core network. 
Each NS is an independent virtualized end-to-end network instance to ensure end-to-end performance for a supported service group. Individual NS for any specific use case can be designed, created and activated dynamically by orchestrating the network functions on the distributed cloud infrastructure. For telesurgery, since multi-modal sensory data are generated, it might need to create individual slices for 3D video, haptic data and physiological data to better fulfil the QoS of each type of sensory data. Virtualization allows instantiation of network functions at optimal locations in the forwarding graph to shorten the latency. The centralized SDN controller facilitates the low-latency forwarding path discovery and reliable multipath routing. Additionally, network coding (NC) combined with SDN can significantly reduce the latency and improve reliability exploiting the recoding and sliding window features of NC. The isolation among NSs can improve security and privacy. 

Furthermore, the real-time traffic monitoring in the centralized controller in SDN can flexibly adjust bandwidth for the traffic fluctuation. The dynamic programmability and control in SDN, data analytics and machine learning, make it feasible to dynamically programming of network slices to optimize the communication performance and user experience. For example, with network data analytics to predict network performance variation, network slices can be dynamically programmed to smoothly ensure the end-to-end performance.


Comparing with the expensive private network, network slicing can significantly improve the network utilization and offer cost-effective communication services. When the communication cost becomes trivial in the overall surgery cost, it will become a good business case for both healthcare industry and network operator, thereby propelling telesurgery to be adopted as medical routine basis.

\subsection{Artificial Intelligence}\label{subsec:AI}
Artificial Intelligence, in particular, convolutional neural network and deep learning, will play an important role in telesurgery. Besides offering diverse value added services, e.g., abnormality diagnosis and positioning, dissection strategy, various techniques based on AI can be designed and developed to improve safety, security, accuracy and contribute to the stability of haptic control and reduce the congestion in the core network, thereby improve the overall latency and reliability. 

The software-generated force and position signals can be realized and applied to human operators to improve the safety and accuracy. For example, a virtual ``wall" can be placed around anatomical structure to keep the surgical instrument from contacting it~\cite{Okamura2009}. The predictive and interpolative/extrapolative modules can be deployed at the edge cloud to reduce the latency and packet loss impact on haptic control instability~\cite{Aijaz2016J}. Moreover, the AI engines can simultaneously process the 3D video stream and the on-going manipulation commands, and monitor the latency variation in the network. If the AI engines foresee the risks of a manipulation error, e.g., the robotic arm will cut an artery by mistake due to the delayed display of video stream, some software can generate a force to postpone or hinder the execution of manipulation meanwhile give warnings to the surgeon. This will allow the surgeon to hold on the motion, thereby avoid the operation errors caused by the delayed arrival of the sensory data feedback. 

Additionally, the AI engine can monitor manipulation commands and haptic feedback to perform real-time verification of the manipulation. The historically collected surgical motions can verify the manipulation commands. In this way, any robotic malfunction or unauthorized operation task due to security attack will be timely and reliably detected.

\section{Concluding Remarks}
Robotic telesurgery will offer a paradigm shift in healthcare and expand telemedicine domain. It brings unprecedented opportunities as well as crucial challenges. This paper reviews the state-of-art robotic telesurgical system, identifies the limiting factors, and summarizes the communication QoS requirements of the multi-modal sensory data and point out the relevant research challenges, open issues and the enabling technologies in the 5G communication system. It is clear that the 5G communication system will play a critical role in tackling these challenges and transforming healthcare.

To realize telesurgery and even make it commercialized, it needs not only to overcome the technical challenges but also to be supported by a sustainable business model and legal regulations. Envisioning the roadmap of telesurgery development and its wide adoption, it is expected to be firstly applied in telementoring scenarios using the dual consoles, namely the experienced surgeon remotely conducts the most complicated tasks in the surgery. Such application scenarios have a good business case as those skills are not commonly available; furthermore, it takes shorter time consuming less communication resource and is cost effective.

\ifCLASSOPTIONcompsoc
  \section*{Acknowledgments}
\else
  \section*{Acknowledgment}
\fi

The authors would like to thank J{\o}rgen Bjerggaard Jensen, Clinical Professor in Urology at Aarhus University with special focus on bladder cancer and robotic surgery, for his experience in working with robotic surgery.

\ifCLASSOPTIONcaptionsoff
  \newpage
\fi



%

\bibliographystyle{IEEEtran}

\end{document}